\documentclass[prl,superscriptaddress,letterpaper,twocolumn,natbib,nofootinbib,lengthcheck,nopreprintnumbers,showpacs]{revtex4-1}
\usepackage{aas_macros}
\usepackage{natbib}
\usepackage{amsmath,amssymb}
\usepackage{graphicx} 
\usepackage{rotating}
\usepackage{amsmath}
\usepackage[caption=false]{subfig}
\usepackage{bm}
\usepackage{tabularx}
\usepackage[usenames]{color}
\usepackage{multirow}
\usepackage{xspace}
\usepackage[plainpages=false, colorlinks=true, anchorcolor=blue, linkcolor=blue, citecolor=blue, bookmarks=false]{hyperref}
\newcolumntype{b}{X}
\newcolumntype{s}{>{\hsize=.5\hsize}X}

\def\be{\begin{equation}}
\def\ee{\end{equation}}
\newcommand{\bb}{\begin{bmatrix}}
\newcommand{\eb}{\end{bmatrix}}

\def\bea{\begin{align}}
\def\eea{\end{align}}

\def\be{\begin{equation}}
\def\en{\end{equation}}
\def\bea{\begin{eqnarray}}
\def\ena{\end{eqnarray}}

\definecolor{ao(english)}{rgb}{0.0, 0.5, 0.0}

\begin{document}

\title[Gaussian process models for PTA data analysis]{Constraints On The Dynamical Environments Of \\Supermassive Black-hole Binaries Using Pulsar-timing Arrays}

\author{Stephen R. Taylor}
\email[]{Stephen.R.Taylor@jpl.nasa.gov}
\affiliation{Jet Propulsion Laboratory, California Institute of Technology, 4800 Oak Grove Drive, Pasadena, CA 91106, USA}
\author{Joseph Simon}
\affiliation{Center for Gravitation, Cosmology and Astrophysics, University of Wisconsin Milwaukee, PO Box 413, Milwaukee WI 53201, USA}
\author{Laura Sampson}
\affiliation{Center for Interdisciplinary Exploration and Research in Astrophysics (CIERA) and Department of Physics and Astronomy, Northwestern University, 2131 Tech Drive, Evanston, IL 60208, USA}

\date{\today}

\begin{abstract}

We introduce a technique for gravitational-wave analysis, where Gaussian process regression is used to emulate the strain spectrum of a stochastic background by training on population-synthesis simulations. This leads to direct Bayesian inference on astrophysical parameters. For PTAs specifically, we interpolate over the parameter space of supermassive black-hole binary environments, including $3$-body stellar scattering, and evolving orbital eccentricity. We illustrate our approach on mock data, and assess the prospects for inference with data similar to the NANOGrav $9$-yr data release. 

\end{abstract}


\maketitle

\section{Introduction}
\label{sec:intro}
Expensive computer simulations are a common problem in astrophysics. Complicated studies, from the details of exoplanet formation to cosmological galaxy evolution, require numerical simulations that can take hours to months~\cite{Vogelsberger:2014kha,Fitts:2016usl,2016arXiv160900437S}. Each simulation initiates from a set of physical assumptions or input parameters. Bayesian estimation of these parameters using experimental data requires a full exploration of possible values, i.e. simulations at each point in parameter space, becoming prohibitively computationally expensive. In this Letter, we describe a method to avoid this problem using the technique of Gaussian process (GP) regression \citep{rw06}. We demonstrate this method with astrophysical inference of supermassive black-hole binary (SMBHB) environments using the nanohertz gravitational waves (GWs) that will soon be detected by pulsar timing arrays (PTAs) \citep{fb90}.

The detection of the stochastic background of GWs from the mergers of SMBHBs throughout our universe will likely occur within the next $\sim 10$ years~\cite{tv+15,sejr13}.  
As discussed in \citet{arz+15} (N16), and \citet{scm15}, the details of the GW background are sensitive to the physics that drives mergers. This can include binary eccentricity, $3$-body scattering encounters with stars in the galactic nucleus, accretion disk dynamics, and much more --- all of which are theorized solutions to the \textit{final parsec problem} \citep{mm03}, which arises because dynamical friction in post-merger galaxy remnants is insufficient to drive SMBHs to milliparsec orbital separations \cite{bbr80}. It is at these separations that GW emission dominates binary evolution, and ultimately drives the binaries to merger. Since we do not observe multiple SMBHs in the cores of galaxies \cite{bs11, wgl+17}, we expect that most galactic mergers lead to BH mergers, and so some confluence of mechanisms must solve the final parsec problem \cite{kfm+16}.

In this Letter we show that it is possible to perform full Bayesian inference on the detailed physics of SMBHB mergers by training a GP on a small set of simulations, initialized on a grid in input parameter space. We then use the GP as a prior on the shape of the GW spectrum in PTA data. This prior is a function of astrophysical parameters, so our analysis directly samples the posterior distributions of these parameters.

Analytic spectral models exist to encapsulate the influences of stars/gas \citep{scm15} or eccentricity \emph{alone} \citep{hmgt15, csd16} on the strain spectrum --- this type of analysis was performed in N16. Stellar scattering, though, impacts SMBHB eccentricity evolution~\citep{rs12}, and so we must model their combined influence. GP regression gives us the first PTA technique capable of constraining the \emph{combined} dynamical influences of a dense galactic-center stellar distribution and evolving SMBHB eccentricity.

\section{Methodology}
\label{sec:train_gp}

GP regression is a powerful interpolation scheme which treats (noisy) data as a random draw from a multivariate Gaussian process with a mean vector and covariance function. We use the initial training data to learn the GP's covariance structure, after which we make predictions about the outcome of hypothetical experiments between the training points (interpolation), and beyond them (extrapolation). Our assumption is that the data do not exhibit pathological discontinuities that would be poorly captured by a smooth covariance (kernel) function. GP regression allows us to interpolate and extrapolate beyond the initial training data, and additionally provides the uncertainty in the prediction which we can propagate forward to our final statistics. It has recently been investigated in a ground-based GW context to marginalize over the uncertainties between expensive numerical relativity waveforms and cheaper post-Newtonian waveforms \citep{mg14,gm15,mbcg16}, but has not been used in GW data-analysis to emulate the statistical properties of astrophysical populations.

Our training data are GW spectra built from synthesized populations of SMBHBs, evolved from different initial conditions. In this proof-of-concept study we restrict our attention to post--dynamical-friction binary eccentricity and stellar densities in the galactic core. Our technique can be trivially expanded to include a wider range of galactic-center environmental influences and assembly-history processes. 

We build and train a GP capable of predicting the shape of the strain spectrum when binary populations have (i) non-zero eccentricity; (ii) significant orbital evolution driven by stellar hardening; (iii) all of the above. Our procedure is as follows:

\begin{enumerate}\itemsep0em 
\item \textbf{Simulate training data:} Build a bank of SMBHB populations by initializing simulations with different binary eccentricities and environments.  
\newline\textit{[Computationally expensive]}. 
\item \textbf{Train the Gaussian-process model:} Model the strain distribution over population realizations as Gaussian with a mean and standard error.  
This noisy data is used to train a GP and optimize its kernel hyper-parameters.\newline\textit{[Computationally cheap]}.
\item \textbf{Analyze PTA data with Gaussian-process model:} The trained GP predicts the shape of the strain spectrum for our GW analysis. 
\newline\textit{[Computationally cheap]}.
\end{enumerate}

We now briefly discuss steps $1$ and $2$. Step $3$ is briefly described in the results section.

\subsection{Population synthesis}
\label{sec:popsynth}
The orbital eccentricity and environmental couplings of a population of SMBHBs do not directly impact the merger-rate density. 
They effect the evolution of the binaries and the frequency distribution of the characteristic strain emitted by each source, $h_{c} (f_{gw})$. Thus, all populations share a common binary merger-rate density prescription like that described in \citet{ss16}, which utilizes observations of galaxy stellar mass functions (GSMFs), galaxy close-pair fractions (${\it f}_{\rm pair}$), and black hole-host galaxy relations ($M-M_{\rm bulge}$) to infer the rate density. To ease computational burden, a single measurement of each observable was used: GSMF from \citet{iml+13}, ${\it f}_{\rm pair}$ from \citet{rdd+14}, and $M$$-$$M_{\rm bulge}$ from \citet{mm13}. For each realization, the value of each observable was randomly drawn from a Guassian with widths equal to the cited one-sigma uncertainty regions, this method propagates observational uncertainties through model inference. 

A SMBHB's orbital evolution from interactions with a fixed, isotropic, unbound, cuspy stellar background are described in \citet{rws+14}, which draws extensively on numerical simulations from \citet{q96}, \citet{shm06}, and \citet{s10}. As in \citet{rws+14}, we anchor all binary evolution to a starting frequency of ${\it f}_{gw} = 10^{-12}$ Hz, which is sufficiently low that any GW emission is outside PTA sensitivity ranges. The initial eccentricity parameter, $e_{0}$, is set at this frequency, and is evolved across the PTA band in accordance with whichever mechanism (stars or GWs) dominates the orbital evolution.  

To generate a population, we draw a finite number of sources whose binary parameters match the merger-rate density for each realization, with eccentricities that have been evolved according to the prescribed environmental conditions $\{ e_{0}, \rho \}$, where $\rho$ is the mass density of stars at the gravitational influence radius of the binary. We assume that all binaries have the same initial eccentricity and are embedded in a stellar distribution with the same density. In future work we will study how varying the distribution of environmental conditions across sources effects the performance of our method. 

\label{sec:buildstrain}
\begin{figure}[!t]
\begin{center}
\includegraphics[width=0.5\textwidth]{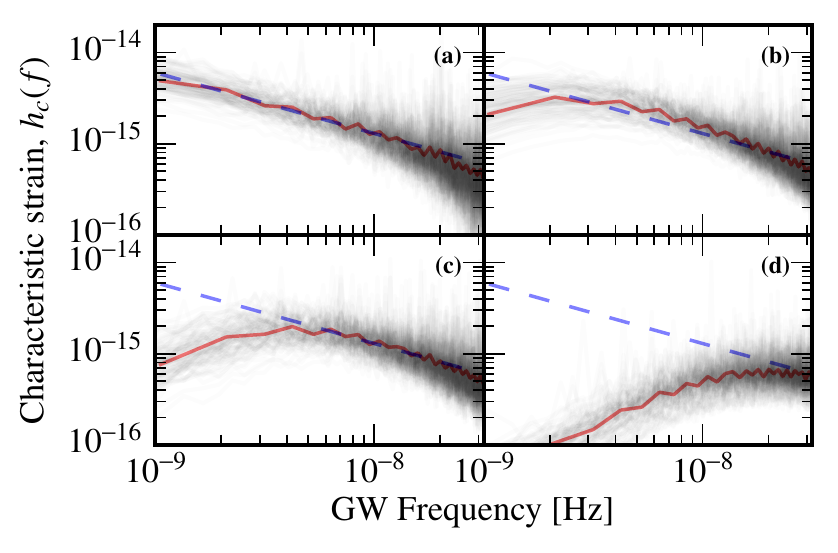}
\end{center}
\caption{Characteristic strain spectra for binary populations under various astrophysical conditions. Grey lines indicate single population realizations, while red is the mean over realizations. Dashed blue shows a $\propto f^{-2/3}$ strain spectrum for reference. (a) $\{e_0 = 0, \rho = 10\, M_\odot \mathrm{pc}^{-3}\}$; (b) $\{e_0 = 0.95, \rho = 10\, M_\odot \mathrm{pc}^{-3}\}$; (c) $\{e_0 = 0, \rho = 10^4\, M_\odot \mathrm{pc}^{-3}\}$; (d) $\{e_0 = 0.95, \rho = 10^4\, M_\odot \mathrm{pc}^{-3}\}$}
\label{fig:starsecc_extremespectra}
\end{figure}

Almost all ($>99\%$) of the GW strain in the PTA sensitivity band ($ 10^{-10} < {\it f}_{gw} < 10^{-7}$ Hz) comes from less than $2\times 10^5$ sources. These sources are saved for each realization. The characteristic strain spectrum of the GW background, $h_{c}(f)$, is built up as the quadrature sum of the strain from each source. The eccentricity of each source determines how the strain is distributed across frequency, with quasi-circular sources emitting mostly in the $n=2$ harmonic of the orbital frequency, while more eccentric sources emit across tens to hundreds of harmonics \citep{pm63}. 
Figure \ref{fig:starsecc_extremespectra} shows the characteristic strain spectra at the four corners of the combined parameter space $\{ e_{0}, \rho \}$ explored in this work.

\subsection{Training the Gaussian process}
\label{sec:traingp_explain}
Our work uses the \texttt{George} GP regression library \citep{hodlr}.\footnote{\href{http://dan.iel.fm/george/current/}{http://dan.iel.fm/george/current/}} We train a separate GP at each sampling frequency of the GW spectrum, from a base frequency of $f=1/T$ up in increments of $1/T$ for a pre-defined number of modes, where $T$ is the span between the maximum and minimum TOAs in the PTA dataset.\footnote{For realistic datasets most of the spectral information is confined to the lowest few frequencies, but we model at least $20$.} While we assume that variation of the strain across astrophysical parameter space is smooth, we do not enforce that it is smooth across frequencies.

We construct the strain spectra for $100$ population realizations at each of $14\times13=182$ combinations in $\{e_0,\rho\}$ parameter space, where $e_0\in\mathcal{U}[0.0,0.95]$ and $\log_{10}(\rho / M_\odot\mathrm{pc}^{-3})\in\mathcal{U}[1,4]$. We then form a data vector at each GW frequency, corresponding to the base-$10$ logarithm of the mean squared characteristic-strain, $\log_{10}\langle h_c^2\rangle$, for every set of dynamical conditions. Each data point has an associated uncertainty corresponding to the log-space standard deviation of the squared strain over the $100$ population realizations. 

To characterize the GP, we choose a \textit{Squared Exponential} (SE) kernel function with a flat, diagonal metric. This stationary, infinitely differentiable function has tunable length-scale parameters which set the correlation between data points across parameter space, and in our case has the form
\begin{equation}
k(\vec{x},\vec{x'}) = \sigma_k^2 \exp{ \left(- (x_i-x'_i)^2 / 2\sigma_i^2 - (x_j-x'_j)^2 / 2\sigma_j^2 \right) },
\end{equation}
where $\vec{x}$, $\vec{x'}$ are two coordinates in our 2-d $\{e_0,\log_{10}\rho\}$ parameter space, each dimension of the input coordinate can have a separate variance $\sigma_i^2$, $\sigma_j^2$, and the kernel has an overall variance scaling, $\sigma_k^2$.\footnote{We investigated the impact of kernel choices on our results, and found that both \textit{Mat\'ern $3/2$} and \textit{Mat\'ern $5/2$} kernels \citep{stein2012interpolation} give comparable posterior recoveries with respect to the SE kernel.}

\begin{figure}[!t]
\begin{center}
\includegraphics[width=0.5\textwidth]{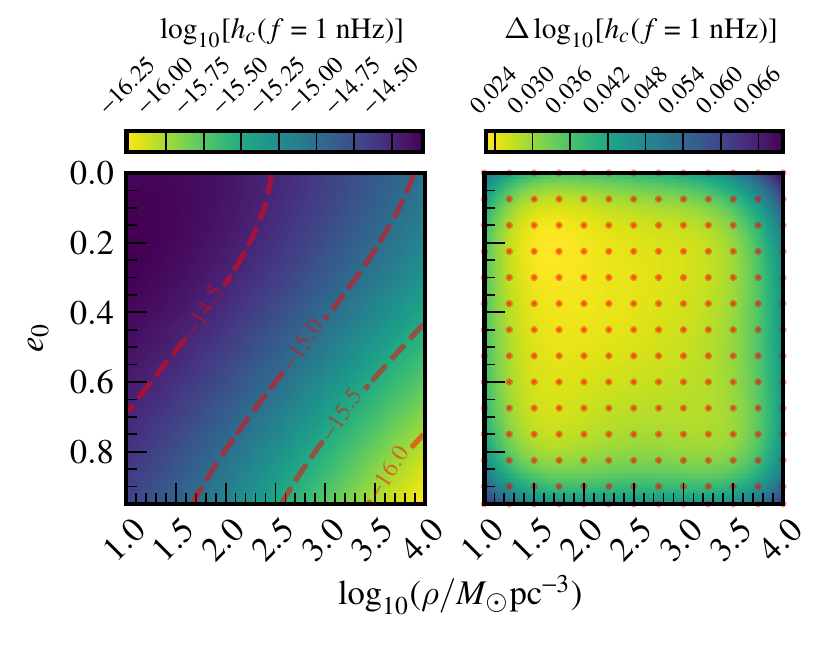}
\end{center}
\caption{Mean (\textbf{left}) and standard deviation (\textbf{right}) of a GP trained on the strain spectrum at $f~=~1/(30\,\mathrm{yr})~\sim~1\,\mathrm{nHz}$. The training spectra are constructed from $100$ synthesized SMBHB populations at each of $182$ different points in $\{e_0,\rho\}$ space (right, red points), with kernel length scales of $10$ and $0.83$, respectively. The uncertainty in the GP prediction is dominated by the intrinsic scatter in the simulations.}
\label{fig:gp4ptas_starsecc_trained}
\end{figure}

The data vector at each frequency is used to train the GP by mapping out the posterior distribution of the kernel function parameters. We set the kernel parameters to their \textit{maximum-a-posteriori} values. An example is shown in Fig.\ \ref{fig:gp4ptas_starsecc_trained} for a GW frequency of $1/(30\,\mathrm{yr})$, where the left panel shows the coordinates of the training data (red points) along with the prediction of the trained-GP. The right panel shows the uncertainty in this prediction.

\begin{figure}[!t]
\begin{center}
\includegraphics[width=0.5\textwidth]{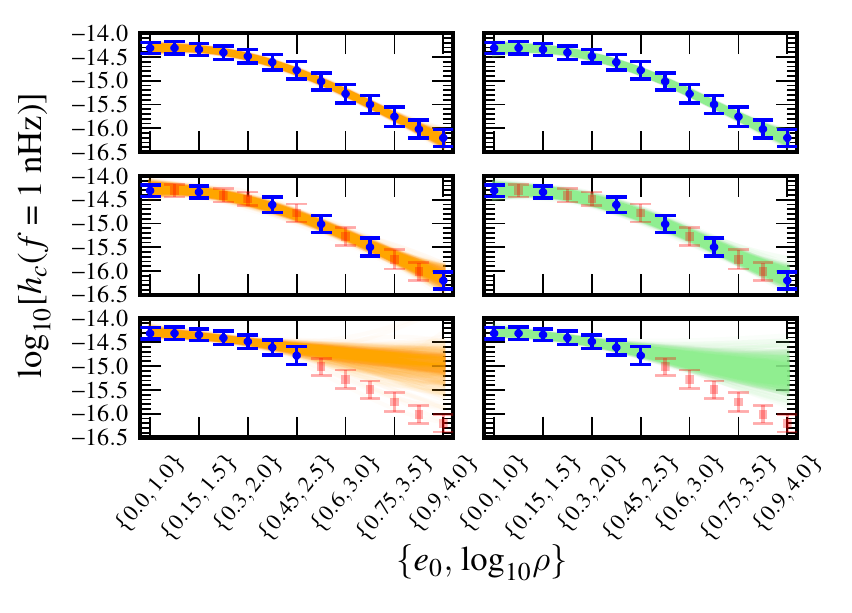}
\end{center}
\caption{Interpolation and extrapolation fidelity of the GP, trained on only the blue points. These points with uncertainties are combinations of $\{e_0,\log_{10}\rho\}$ along the downwards left-right diagonal of Fig.\ \ref{fig:gp4ptas_starsecc_trained}'s left panel, where the strain is varying fastest. Faded red squares show data excluded from the training. Colored regions are random draws from the trained GP. The \textbf{left column} shows $10$ random draws from the GP with each of $50$ different kernel parameter values drawn from their posterior distribution. The \textbf{right column} is similar, except all $500$ draws use the \textit{maximum-a-posteriori} kernel parameters.} 
\label{fig:gp4ptas_starsecc_testremove}
\end{figure}

Figure \ref{fig:gp4ptas_starsecc_testremove} shows tests of the interpolation and extrapolation fidelity of the GP at $f=1/(30\,\mathrm{yr})$, achieved by removing some points from the training data. Interpolation performs very well, but extrapolation is poor, so we do not use our GP model outside of the boundaries of the original training data.\footnote{Performance is graded on how far the removed points depart from the GP prediction, in units of the GP uncertainty. For interpolation all removed points lie within the $1$-$\sigma$ region, while for extrapolation the removed points start at $\sim 1$-$\sigma$ away, and depart farther from the GP prediction as we move farther outside the boundaries of the training set.} Figure \ref{fig:gp4ptas_starsecc_testremove} also shows that there is no noticeable difference in performance when fixing the kernel parameters to their \textit{maximum-a-posteriori} values compared to sampling from their posterior distribution. As we expand our GP model to more astrophysical influences in the future, we will monitor this to assess whether kernel parameter sampling is also needed as part of our hierarchical Bayesian pulsar-timing model.

Finally, for this study we are interested in the influence of dynamical processes on the \emph{shape} of the GW spectrum, and not on the overall amplitude implied by merger-rate density assumptions. Hence, before training we find the mean amplitude of the $\{e_0=0,\rho=10\,M_\odot\mathrm{pc}^{-3}\}$ simulations at $f=1/\mathrm{yr}$, then divide all spectra by this amplitude. 

\section{Results}
\label{sec:results}
Our GP model fits into existing Bayesian analysis schemes as an extension of the hierarchical model. Current schemes expand the time-domain GWB signal onto a Fourier basis, then marginalize over the Fourier amplitudes with a Gaussian prior \citep{l+13,vv14}. This prior is parametrized to have a variance that obeys e.g.\ $(i)$ a power-law spectrum with amplitude and spectral index parameters; $(ii)$ a turnover model with three `shape' parameters \citep{scm15}; $(iii)$ a free spectrum with a separate amplitude parameter per GW frequency. 

We use and extend model $(iii)$ by placing a Gaussian prior on the power spectrum at each GW frequency with a mean and standard deviation from the GP prediction. We numerically sample the power spectrum at each frequency ($n$ parameters) and the hyper-parameters of the GP model ($\{A,e_0,\log_{10}\rho\}$), for a GWB model space with $n+3$ dimensions. To ensure efficient numerical sampling, we perform a coordinate transformation such that the power spectrum at each frequency is sampled as a zero-mean unit-variance Gaussian~\citep{n03}. The GP method is implemented within the NX01 Bayesian PTA package \citep{nx01},\footnote{\href{https://github.com/stevertaylor/NX01}{https://github.com/stevertaylor/NX01}.} and we use a parallel-tempering MCMC sampler with several bespoke proposal schemes.\footnote{\href{https://github.com/jellis18/PTMCMCSampler}{https://github.com/jellis18/PTMCMCSampler}.}

We test the performance of our method in both the presence and absence of a detectable GW background. We use two types of data sets: $(1)$ a simplified dataset consisting of an array of $18$ pulsars whose positions and statistical noise properties match those used in N16. These pulsars are concurrently timed for $30$ years with a total of $390$ TOAs in each. $(2)$ A realistic dataset consisting of an $18$-pulsar array that emulates the N16 PTA in every way, with the same noise properties and an identical observation schedule over a $\sim 9$ year baseline \citep{arz+15b}. We calibrate this dataset to give the reported $95\%$ upper limit of $1.5\times10^{-15}$ on the amplitude of an $f^{-2/3}$ GWB power-spectrum at $f=1/\mathrm{yr}$. This required us to inflate the TOA uncertainties by a factor of $1.4$ across all pulsars.

\textit{Type-$(1)$---} We create two type-$(1)$ datasets with differing injected GWB spectra. First, we create $100$ new population realizations with dynamical conditions $\{e_0~=~0.65,~\log_{10}(\rho / M_\odot\mathrm{pc}^{-3})~=~3.35\}$, and inject the mean characteristic-strain spectrum. This combination of conditions was chosen to provide a strain-spectrum turnover within the frequency coverage of a $30$ year data span, so that both dynamical properties could be constrained. We also deliberately excluded this simulation from the training data. Second, we inject a pure $f^{-2/3}$ power-law strain spectrum. In both cases, the overall amplitude scaling is set such that the Bayes factor for spatial correlations $\gtrsim 10^5$, because any inaccuracies in the GP model are best illuminated with parameter estimation in high signal-to-noise scenarios where the posterior uncertainty is small and modeling inaccuracies will show as systematic offsets. 

\begin{figure}[!t]
\begin{center}
\includegraphics[width=0.5\textwidth]{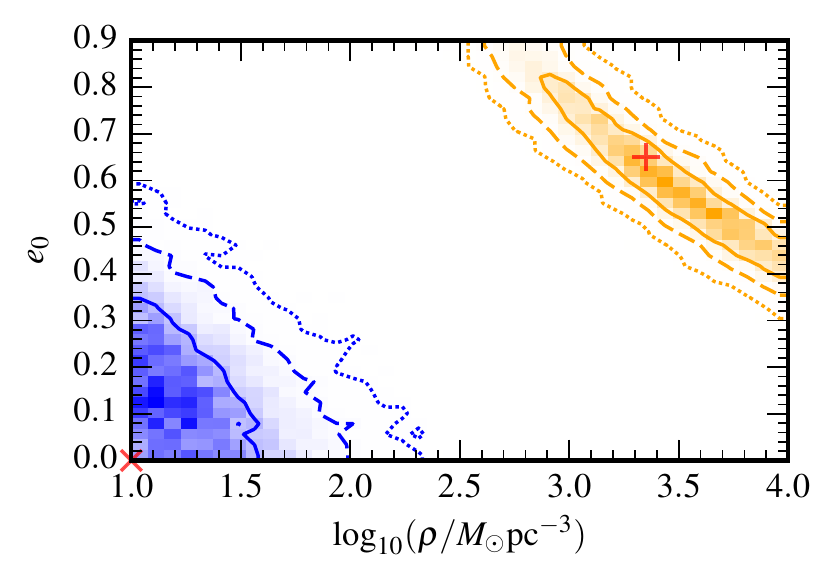}
\end{center}
\caption{Type-$(1)$ datasets (see text) are created with injected GWB signals whose spectra are influenced by the dynamical environments indicated by $+$ and $\times$. The recovered posterior distributions from employing our GP search model are shown as orange and blue density regions respectively, with associated $68\%$ (solid), $95\%$ (dashed), and $99.7\%$ (dotted) credible intervals.}
\label{fig:two_sim_results}
\end{figure}

The white noise was fixed at the level of the TOA uncertainties, while the GWB signal and per-pulsar red-noise parameter priors were uninformative. 
The resulting $2$-d marginalized posterior distributions in $\{e_0,\rho\}$ space are shown in Fig.\ \ref{fig:two_sim_results}, where we see that the injected values lie within the $68\%$ credible regions in both cases. The parameter degeneracy arises because a population of highly-eccentric binaries in a moderately-dense stellar environment can give a strain spectrum turnover at a similar frequency as a lower eccentricity population in a more dense stellar environment. Similar physical parameter degeneracies were observed in \citet{cms+16}.

\begin{figure}[!t]
\begin{center}
\includegraphics[width=0.5\textwidth]{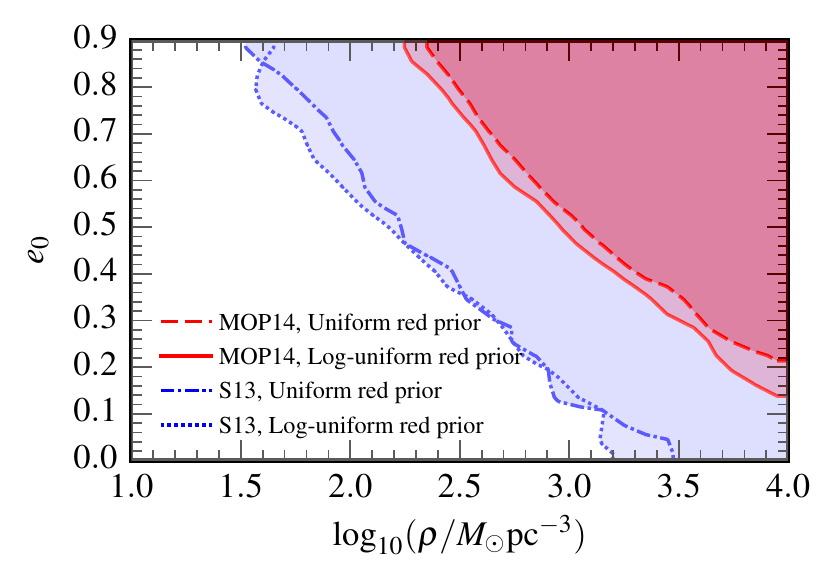}
\end{center}
\caption{Astrophysical inference on an emulated NANOGrav $9$-yr dataset. The GWB amplitude at high frequencies is anchored to the MOP14 and the S13 priors. The lines indicate the boundaries of the $68\%$ credible regions under different prior assumptions. See the text for details and discussion.}
\label{fig:nano9yr_results}
\end{figure}

\textit{Type-$(2)$---} No GWB signal is injected in this dataset, since N16 reported no evidence for GWs. Instead, we perform astrophysical inference on this fake dataset as was performed on the true dataset. The parameter priors were as in the Type-$(1)$ datasets above. 
Instead of searching for a GWB amplitude with a prior that is uniform in log-space, we anchor the amplitude with priors that agree with two different astrophysical predictions at high GW frequencies. These predicted amplitude distributions derive from \citet{mop14} (MOP14), and \citet{s13} (S13), where the assumptions that led to them are explained in the relevant papers, and contrasted in N16. They respectively constitute an optimistic ($\log_{10}A=-14.4\pm0.26$) and a median ($\log_{10}A=-15.0\pm0.2$) prediction of the GWB amplitude at high frequencies. 

The resulting $2$-d marginalized posterior distributions in $\{e_0,\rho\}$ space are shown in Fig.\ \ref{fig:nano9yr_results} for the different prior assumptions. Under the assumptions of the MOP14 and S13 amplitude priors, the emulated NANOGrav $9$-yr data favors eccentric SMBHBs that are coupled to dense stellar distributions. This is because an $f^{-2/3}$ power-law spectrum with MOP14 or S13 amplitudes would produce a GWB signal in excess of current constraints at low frequencies. To be consistent with the absence of a GW signal, dynamical influences must attenuate the low-frequency GW spectrum. Because it is larger, the MOP14 prior necessitates stronger spectral attenuation and thus more extreme dynamical conditions. With the more recent \citet{s16} amplitude prior (centered on $\log_{10}A=-15.4$) the posterior distribution in $e_0$ and $\log_{10}\rho$ would be flat and uninformative. The choice of red-noise prior has a minimal influence. 

\section{Discussion}
\label{sec:conclusion}
We have introduced a powerful new method for GW data analysis, where a model for the stochastic GW strain spectrum is built entirely through Gaussian-process emulation of population synthesis simulations. This is the first PTA method capable of searching for the \emph{combined} SMBHB dynamical influences of stellar scattering encounters from the loss-cone and evolving binary eccentricity. These effects are important for understanding SMBHB dynamics \cite{scqg13}, but are poorly constrained by current observations. PTAs offer a more direct measurement of sub-parsec binary evolution than is currently available. Further, our technique can be trivially expanded to incorporate additional effects, such as accretion from a circumbinary disk, different $M$$-$$\sigma$ relations, or dynamical-friction timescales.

Our results for a dataset that precisely emulates the sensitivity of the NANOGrav $9$-year dataset indicate that the MOP14 and S13 GWB amplitude predictions mildly favor a population of eccentric SMBHBs evolving within dense stellar distributions. The true dataset gives quantitatively similar results, and will be explored further in a forthcoming analysis \citep{arz+17}.

Beyond pulsar-timing analysis, this method can be adapted for LIGO or LISA population inference. Current schemes perform demographic analysis to recover the distributions of compact-system properties with either a parametric function \citep{acl12,f+11}, or using a histogram with bin heights constrained by a GP prior \citep{s16,f+11}. Linking these distributions back to progenitor properties or evolutionary channels has so far only been performed for discrete population synthesis simulations \citep{sos15}. With our technique, one could use the histogram model with a GP prior trained on a set of population synthesis simulations. This would allow sampling of the continuous posterior distribution of progenitor properties, such as metallicity or the efficiency of common-envelope hardening. We will investigate this in future work.

A notebook that allows the user to read in simulations and construct their own GP model is available at \href{https://github.com/stevertaylor/gw_pta_emulator}{https://github.com/stevertaylor/gw\_pta\_emulator}.

\begin{acknowledgments}
\textit{Acknowledgments---} We thank the anonymous referees for their thorough review and comments that improved this manuscript. We are also indebted to our NANOGrav and EPTA colleagues for their insights and helpful suggestions. We thank Christopher Moore and Jonathan Gair for useful comments on Gaussian process regression. SRT was partly supported by appointment to the NASA Postdoctoral Program at the Jet Propulsion Laboratory, administered by Oak Ridge Associated Universities and the Universities Space Research Association through a contract with NASA. JS was partly supported by a Wisconsin Space Grant Consortium Graduate Fellowship. LMS was supported by a CIERA Fellowship at Northwestern University. This work was supported in part by National Science Foundation Grant No. PHYS-1066293 and by the hospitality of the Aspen Center for Physics. A majority of the computational work was performed on the Nemo cluster at UWM supported by NSF grant No. 0923409. The research was partially carried out at the Jet Propulsion Laboratory, California Institute of Technology, under a contract with the National Aeronautics and Space Administration. \copyright\,~2016. All rights reserved.
\end{acknowledgments}

\bibliography{apjjabb,bib}

\end{document}